\documentclass[final]{IEEEtran}
\usepackage{cite}
\usepackage{threeparttable}
\usepackage{graphicx}
\usepackage{picinpar}
\usepackage[cmex10]{amsmath}
\usepackage{amsmath,amsfonts,amssymb}
\usepackage{subfigure}
\usepackage{enumerate}
\usepackage{color}

\usepackage{cite}
\usepackage{threeparttable}
\usepackage{graphicx}
\usepackage{picinpar}
\usepackage[cmex10]{amsmath}
\usepackage{amsmath,amsfonts,amssymb}
\usepackage{subfigure}
\usepackage{algorithm}
\usepackage{algorithmic}
\usepackage{stfloats}
\usepackage{bm}
\usepackage{amsthm}
\usepackage{changebar}
\usepackage{algorithm}
\usepackage{algorithmic}
\usepackage{stfloats}
\usepackage{bm}
\usepackage{mathrsfs}

\begin{document}
\newtheorem{lemma}{Lemma}
\newtheorem{corol}{Corollary}
\newtheorem{theorem}{Theorem}
\newtheorem{proposition}{Proposition}
\newtheorem{definition}{Definition}
\newcommand{\e}{\begin{equation}}
\newcommand{\ee}{\end{equation}}
\newcommand{\eqn}{\begin{eqnarray}}
\newcommand{\eeqn}{\end{eqnarray}}

\title{Compressive Sensing Based Multi-User Detector for the
  Large-Scale SM-MIMO Uplink}

\author{\authorblockN{Zhen Gao, Linglong Dai, Zhaocheng Wang,~\IEEEmembership{Senior Member,~IEEE}, \\Sheng Chen,~\IEEEmembership{Fellow,~IEEE}, and Lajos Hanzo,~\IEEEmembership{Fellow,~IEEE}}
\thanks {Z. Gao, L. Dai, and Z. Wang are with Tsinghua National Laboratory for
 Information Science and Technology (TNList), Department of Electronic Engineering,
 Tsinghua University, Beijing 100084, China (E-mails: gaozhen010375@foxmail.com; daill@mail.tsinghua.edu.cn; zcwang@mail.tsinghua.edu.cn).}
\thanks {S. Chen and L. Hanzo are with Electronics and Computer Science, University of Southampton, Southampton SO17 1BJ, U.K. (E-mails: sqc@ecs.soton.ac.uk; lh@ecs.soton.ac.uk), S. Chen is also with King Abdulaziz University, Jeddah 21589, Saudi Arabia.
.}
\thanks {This work was supported in part by the International Science \& Technology Cooperation Program of China (Grant No. 2015DFG12760),
the National Natural Science Foundation of China (Grant Nos. 61571270 and 61201185),  the Beijing Natural Science Foundation (Grant No. 4142027),
and the Foundation of Shenzhen government.}

}

\maketitle

\begin{abstract}
Conventional spatial modulation (SM) is typically considered for
transmission in the downlink of small-scale MIMO systems, where a
single one of a set of say $2^p$ antenna elements (AEs) is activated
for implicitly conveying $p$ bits. By contrast, inspired by the
compelling benefits of large-scale MIMO (LS-MIMO) systems, here we
propose a LS-SM-MIMO scheme for the uplink (UL), where each user having
multiple AEs but only a single radio frequency (RF) chain invokes SM
for increasing the UL-throughput. At the same time, by relying on hundreds of AEs but a small number of RF
chains, the base station (BS) can simultaneously serve multiple users whilst reducing the
power consumption. Due to the large number of AEs of the UL-users and the comparably small number of RF chains at the
BS, the UL multi-user signal detection becomes a challenging
large-scale under-determined problem. To solve this problem, we propose
a joint SM transmission scheme and a carefully designed structured compressive sensing
(SCS)-based multi-user detector (MUD) to be used at the users and BS, respectively. Additionally, the cyclic-prefix single-carrier (CPSC) is used to combat the multipath channels, and a simple receive AE selection is used for the improved performance over correlated Rayleigh-fading MIMO channels. We demonstrate that the aggregate SM
signal consisting of multiple UL-users' SM signals in one CPSC block appears the distributed sparsity. Moreover,
due to the joint SM transmission scheme, aggregate SM signals in the same transmission group exhibit the group sparsity. By exploiting these intrinsically
sparse features, the proposed SCS-based MUD can reliably detect the resultant SM signals with low complexity. Simulation
results demonstrate that the proposed SCS-based MUD achieves a better signal detection performance than its
counterparts even with higher UL-throughtput.
\end{abstract}


\section{Introduction}\label{S1}

A widely recognized consensus is that the fifth-generation (5G) systems
will be capable of providing significant energy efficiency and system
capacity improvements~\cite{{5Gbe},{LS_MIMO_SM}}. Promising techniques, such as
large-scale MIMO (LS-MIMO) and spatial modulation (SM)-MIMO systems
are considered as potent candidates for 5G~\cite{5Gbe,LS_MIMO_SM,OFDM_SM,LS_MIMO,LS_SM}.
LS-MIMO employing hundreds of antenna elements (AEs) at the
base station (BS) is capable of improving the spectral efficiency by
orders of magnitude, but it suffers from the nonnegligible power
consumption and hardware cost due to one specific RF chain usually
required by every AE~\cite{LS_SM}. By using a reduced number of
RF chains, the emerging SM-MIMO activates part of available AEs
to transmit extra information in the spatial domain, and it has
attracted much attention due to its high energy efficiency and reduced
hardware cost~\cite{LS_SM}. However, conventional SM-MIMO is usually
considered in the downlink of small-scale MIMO systems, and therefore its
achievable capacity is limited.  Individually, both technologies have
their own advantages and drawbacks. By an effective combination of
them together, one can envision the win-win situation. 
SM-MIMO is attractive for LS-MIMO systems, since the reduced number of
required RF chains in SM-MIMO can reduce the power consumption and
hardware cost in conventional LS-MIMO systems. Moveover, hundreds of AEs
used in LS-MIMO can improve the system throughput of
SM-MIMO. Such reciprocity enables LS-MIMO and SM-MIMO to enjoy the
apparent compatibility.

In this paper, we propose a LS-SM-MIMO scheme for intrinsically
amalgamating the compelling benefits of both LS-MIMO and SM-MIMO for the 5G
uplink (UL) over frequency-selective fading channels. In the proposed scheme, each UL-user equipped with
multiple AEs but only a single RF chain invokes SM for increasing the
UL-throughput, and the cyclic-prefix single-carrier (CPSC) transmission scheme is adopted to combat the multipath channels \cite{CPSC}. At the BS, hundreds of AEs but only dozens of
RF chains are employed to simultaneously serve multiple users, and a direct AE selection scheme is used to improve the system performance over correlated Rayleigh-fading MIMO channels at the BS \cite{mide}. The proposed
scheme can be adopted in conventional LS-MIMO as a specific UL-transmission mode for reducing the power consumption, or
altarenatively, for energy- and cost-efficient LS-SM-MIMO, where joint
benefits of efficient AE selection~\cite{mide}, transmit precoding~\cite{precoding}, and channel
estimation~\cite{CE_SM} can be
readily exploited. To sum up, the proposed scheme inherits the advantages of
LS-MIMO and SM-MIMO, while reducing the power consumption and hardware
cost.

A challenging problem in the proposed UL LS-SM-MIMO scheme is how to realize a reliable multi-user detector (MUD) with low complexity. The optimal maximum likelihood (ML) signal detector suffers from the excessive complexity. Conventional sphere decoding detectors cannot be readily used in multi-user scenarios and may still appear the high complexity for LS-SM-MIMO~\cite{SD}. Existing low-complexity linear signal detectors, e.g.,
the minimum mean square error (MMSE)-based signal detector, perform
well for conventional LS-MIMO systems~\cite{LS_MIMO}. However, they are
unsuitable for the proposed LS-SM-MIMO UL-transmission, since the large number of
transmit AEs of the UL-users and the reduced number of receive RF chains at the BS make the UL multi-user signal detection be a large-scale
under-determined/rank-deficient problem. The authors
of~\cite{CS_CL1,CS_CL2,shim} proposed compressive sensing (CS)-based
signal detectors to solve the under-determined signal detection problem
in SM-MIMO systems, but they only considered the single-user small-scale SM-MIMO systems in the downlink.

Against this background, our new contribution is that we exploit the
specific signal structure in the proposed multi-user LS-SM-MIMO UL-transmission, where
each user only activates a single AE in each time slot. Hence the SM
signal of each UL-user is sparse with the sparsity level of one, and the aggregate SM signal consisting of multiple UL-users' SM
signals in one CPSC block exhibits a certain distributed
sparsity, which can be beneficially exploited for improving the signal detection
performance at the BS. Moreover, we propose a joint SM transmission scheme
for the UL-users in conjunction with an appropriately structured
compressive sensing (SCS)-based MUD at the BS. The proposed SCS-based MUD is
specifically tailored to leverage the inherently distributed sparsity
of the aggregate SM signal and the group sparsity of multiple
aggregate SM signals owing to the joint SM transmission scheme for reliable signal detection performance. Our simulation results demonstrate that the
proposed SCS-based MUD is capable of outperforming the conventional
detectors even with higher UL-throughput.

 The rest of the paper is organized as follows. Section~\ref{S2}
 introduces the system model of the proposed LS-SM-MIMO
 scheme. Section~\ref{S3} specifies the proposed joint SM transmission and SCS-based MUD. Section~\ref{S4} provides our simulation results. Section~\ref{S5} concludes this paper.

 Throughout this paper, lower-case and upper-case boldface letters
 denote vectors and matrices, respectively, while $\left( \cdot
 \right)^{\rm T}$, $\left( \cdot \right)^{*}$, $\left( \cdot
 \right)^{\dag}$ and $\lfloor \cdot \rfloor$ denote the transpose,
 conjugate transpose, Moore-Penrose matrix inversion, and the integer
 floor operators, respectively. The $l_0$ and $l_2$ norm operations are given by
 $\|\cdot\|_0$ and $\|\cdot\|_2$, respectively. The support set of the vector $\mathbf{x}$ is denoted by
 ${\rm supp} \{\mathbf{x}\}$, and $\left. \bf{x} \right\rangle_i$
 denotes the $i$th entry of the vector ${\bf{x}}$. Additionally,
 $\left. \mathbf{x} \right|_{\Gamma}$ denotes the entries of
 $\mathbf{x}$ defined in the set $\Gamma$, $\left.\mathbf{\Phi}\right|_{\Gamma}$ denotes the sub-matrix whose
 columns comprise the columns of $\mathbf{\Phi}$ that are defined in
 $\Gamma$, and ${\left. {\bf{\Phi }} \right\rangle _\Gamma }$ denotes the sub-matrix whose
 rows comprise the rows of $\mathbf{\Phi}$ that are defined in
 $\Gamma$. The expectation operator is given by ${\rm E}\{\cdot \}$. $\bmod \left( {x,y} \right) = x - \left\lfloor {x/y} \right\rfloor y$ if $y \neq 0$ and $x - \left\lfloor {x/y} \right\rfloor y \neq 0$, while $\bmod \left( {x,y} \right) = y $ if $y \neq 0$ and $x - \left\lfloor {x/y} \right\rfloor y = 0$.

\section{System Model}\label{S2}
\begin{figure}[tb]
\begin{center}
\includegraphics[width=9cm]{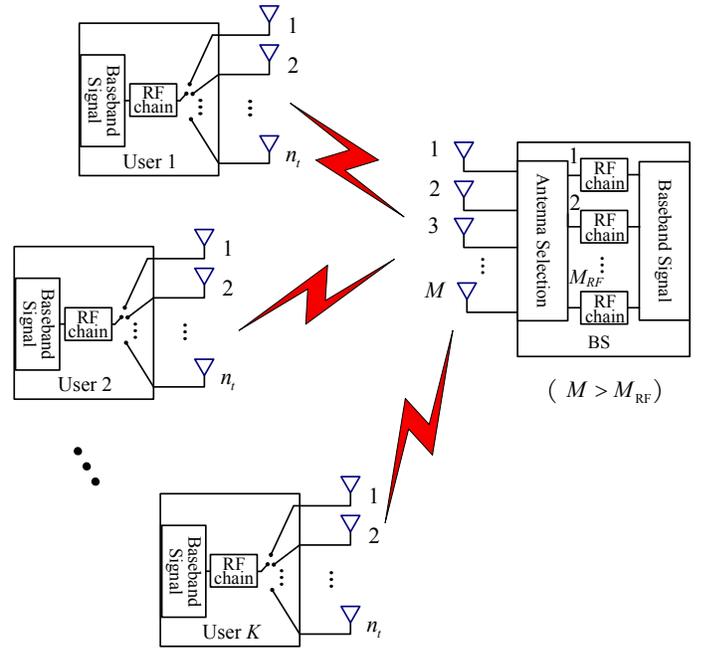}
\end{center}
\caption{In the proposed UL LS-SM-MIMO, the BS is equipped with $M$
  AEs and $M_{\rm{RF}}$ RF chains to simultaneously serve $K$ users,
  where $M \gg M_{\rm{RF}} > K$, and each user is equipped with $n_t >
  1$ AEs and one RF chain. By exploiting the improved degree of
  freedom in the spatial domain, multiple users can simultaneously
  exploit SM for improving the UL-throughput.}
\label{SM_MIMO} 
\end{figure}
 We first introduce the proposed LS-SM-MIMO scheme and then focus our
 attention on the UL-transmission with an emphasis on the multi-user signal detection.

 \subsection{Proposed Multi-User LS-SM-MIMO Scheme}\label{S2.1}

As shown in Fig.~\ref{SM_MIMO}, we consider the proposed
LS-SM-MIMO from both the BS side and the user side.  For conventional
LS-MIMO, the number of AEs employed by the BS is equal to the number
of its RF chains~\cite{LS_MIMO}.  However, the BS in LS-SM-MIMO, as
shown in Fig.~\ref{SM_MIMO}, is equipped with a much smaller number of
RF chains $M_{\rm{RF}}$ than the total number of AEs $M$, i.e., we
have $M_{\rm{RF}} \ll M$. Conventional LS-MIMO systems typically assume
single-antenna users~\cite{LS_MIMO}. By contrast, in the proposed
scheme, each user is equipped with $n_t > 1$ AEs but only a single RF
chain, and SM is adopted for the UL-transmission, where only one of
the available AEs is activated for data transmission.  It has been
shown that the main power consumption and hardware cost of cellular
networks comes from the radio access network~\cite{5Gbe}. Hence, using
a reduced number of expensive RF chains compared to the total number
of AEs at the BS can substantially reduce both the power consumption
and the hardware cost for the operators.  Meanwhile, it is feasible
to incorporate several AEs and a single RF
chain in the handsets.  The resultant increased degrees of freedom in
the spatial domain may then be exploited for improving the
UL-throughput. The proposed scheme can be considered as an optional UL-transmission mode in conventional LS-MIMO systems, where AE
selection schemes may be adopted for beneficially selecting the most
suitable $M_{\rm{RF}}$ AEs at the BS to receive UL SM signals~\cite{mide}. Alternatively, it can also be used for the UL of
LS-SM-MIMO, when advantageously combining transmit precoding, receive AE selection, and
channel estimation~\cite{precoding,mide,CE_SM}.


\subsection{Uplink Multi-User Transmission}\label{S2.2}

 We first consider the generation of SM signals at the users.
 The SM
 signal $\mathbf{x}_k= \mathbf{e}_k s_k$ transmitted by the $k$th user
 in a time slot consists of two parts: the spatial constellation
 symbol $\mathbf{e}_k\in \mathbb{C}^{n_t}$ and the signal
 constellation symbol $s_k\in \mathbb{C}$.  
 $\mathbf{e}_k$ is generated by mapping $\left\lfloor
 \log_2(n_t) \right\rfloor$ bits to the index of the active AE, and
 typically the user terminal employs $n_t=2^p$ AEs, where $p$ is a
 positive integer. Due to only a single RF chain employed at each user, only one entry of $\mathbf{e}_k$ associated with the active AE equals one, and the rest of the entries
 of $\mathbf{e}_k$ are zeros, i.e., we have

\begin{equation}\label{equ:sys_model0} 
{\rm supp}(\mathbf{e}_k)\in \mathbb{A} , ~ \left\| \mathbf{e}_k \right\|_0 = 1 ,
 ~ \left\| \mathbf{e}_k \right\|_2 = 1 ,
\end{equation}
where $\mathbb{A}=\left\{ 1,2, \cdots ,n_t \right\}$ is the spatial
constellation symbol set. The signal constellation symbol comes from
the $L$-ary modulation, i.e., $s_k\in \mathbb{L}$, where $\mathbb{L}$
is the signal constellation symbol set (e.g., 64QAM) of size
$L$. Hence, each UL-user's SM signal carries
the information of $\log_2(L)+\log_{2}(n_t)$ bits per channel use
(bpcu), and the overall UL-throughput is
$K\big(\log_2(L)+\log_2(n_t)\big)$ bpcu.
Users utilize the CPSC scheme to transmit SM signals \cite{CPSC}.
Each CPSC block consists of a cyclic-prefix (CP) with the length of $P-1$
and the following data block with the length of $Q$. Hence the length of each
CPSC block is $Q+P-1$. CP can be used to combat the multipath channels with
the length of $P$. The data block consists of $Q$ successive SM signals.

At the receiver, due to the reduced number of RF chains at the BS,
only $M_{\rm{RF}}$ receive AEs can be exploited to receive signals,
where existing receive AE selection schemes can be adopted to
preselect $M_{\rm{RF}}$ receive AEs for achieving an
improved signal detection performance~\cite{mide}. 
Since the BS can serve $K$ users simultaneously, after the removal of CP,
 the received signal $\mathbf{y}_{q}^{ }\in \mathbb{C}^{M_{\rm{RF}}}$ for $1\le q \le Q$ in the $q$th time slot for one CPSC block, can be
expressed as

 \begin{equation}\label{equ:sys_model2} 
 \begin{array}{l}
{\bf{y}}_q^{ } = \sum\limits_{k = 1}^K {{\bf{y}}_{k,q}^{ }}  + {\bf{w}}_q^{ } = \sum\limits_{p = 0}^{P - 1} {\sum\limits_{k = 1}^K {{{\left. {{{\bf{H}}_{k,p}^{ }}} \right\rangle }_\Theta }} } {{\bf{x}}_{k,\bmod \left( {q - p,Q} \right)}^{ }} + {\bf{w}}_q^{ } \\= \sum\limits_{p = 0}^{P - 1} {\sum\limits_{k = 1}^K {{{{\bf{\tilde H}}}_{k,p}}}^{ } } {{\bf{x}}_{k,\bmod \left( {q - p,Q} \right)}^{ }} + {\bf{w}}_q^{ },
\end{array}
\end{equation}
where ${{\bf{H}}_{k,p}^{ }}\in \mathbb{C}^{M\times n_t}$ is the $k$th user's
MIMO channel matrix for the $p$th multipath component, ${\left. {{\bf{H}}_{k,p}^{ }} \right\rangle _\Theta } = {{{\bf{\tilde H}}}_{k,p}^{ }}\in \mathbb{C}^{M_{\rm{RF}}\times n_t}$, the set $\Theta $ is determined by the used AE selection scheme, elements of $\Theta $ with the cardinality of $M_{\rm{RF}}$ are uniquely selected from the set $\left\{ {1,2, \cdots ,M} \right\}$, $\mathbf{x}_{k,q}^{ }$ has one
nonzero entry, and $\mathbf{w}_q^{ }\in \mathbb{C}^{M_{\rm{RF}}}$ is the
additive white Gaussian noise (AWGN) vector with entries obeying
the independent and identically distributed (i.i.d.) circular symmetric
complex Gaussian distribution with zero mean and a variance of $\sigma^2_w/2$ per
dimension, denoted by $\mathcal{CN}(0,\sigma^2_w)$. ${\bf{H}}_{k,p}^{ } = {\bf{R}}_{\rm{BS}}^{1/2}{\bf{\bar H}}_{k,p}^{ }{\bf{R}}_{{\rm{US}}}^{1/2}$, entries of ${\bf{\bar H}}_{k,p}^{ }$ obey
the i.i.d. $\mathcal{CN}(0,1)$, ${\bf{R}}_{{{\rm{US}}}}^{}$ with the correlation coefficient $\rho_{\rm{US}}$ and ${\bf{R}}_{{{\rm{BS}}}}^{}$ with the correlation coefficient $\rho_{\rm{BS}}$ are correlation matrices at the users and BS, respectively. The element of the $m$th row and $n$th column of ${\rm{R}}_{\rm{BS}}$ (${\rm{R}}_{\rm{US}}$) is $\rho _{{\rm{BS}}}^{\left| {m - n} \right|}$ ($\rho _{{\rm{US}}}^{\left| {m - n} \right|}$). For correlated Rayleigh-fading MIMO channels, the specific $\Theta $ or receive AE selection scheme has an impact on the system performance. In this paper, the direct AE selection scheme is used to maximize the minimum geometric distance between any pair of the selected AEs \cite{mide}. For uniform linear array (ULA), $\Theta  = \left\{ {\varphi  + {m_{{\rm{RF}}}}\left\lfloor {M/{M_{{\rm{RF}}}}} \right\rfloor } \right\}_{{m_{{\rm{RF}}}} = 0}^{{M_{{\rm{RF}}}} - 1}$ with $1\le \varphi \le {\left\lfloor {M/{M_{{\rm{RF}}}}} \right\rfloor }-1$.
(\ref{equ:sys_model2}) can be further
expressed as
\begin{equation}\label{equ:sys_model33} 
 \begin{array}{l}
{\bf{y}}_q^{ } = \sum\limits_{p = 0}^{P - 1} {{{{\bf{\tilde H}}}_p^{ }}{{\bf{x}}_{\bmod \left( {q - p,Q} \right)}^{ }}}  + {\bf{w}}_q^{ },
   \end{array}
\end{equation}
by defining 
${\bf{\tilde H}}_p^{^{ }} = \left[ {{\bf{\tilde H}}_{1,p}^{^{ }}{\bf{\tilde H}}_{2,p}^{^{ }} \cdots {\bf{\tilde H}}_{K,p}^{^{ }}} \right]\in \mathbb{C}^{M_{\rm{RF}}\times (n_t K)}$
 and ${\bf{x}}_q^{ } = {\left[ {{{\left( {{\bf{x}}_{1,q}^{ }} \right)}^{\rm{T}}}{{\left( {{\bf{x}}_{2,q}^{ }} \right)}^{\rm{T}}} \cdots {{\left( {{\bf{x}}_{K,q}^{ }} \right)}^{\rm{T}}}} \right]^{\rm{T}}} \in { \mathbb{C}^{({n_t}K)}}$. By considering $Q$ SM signals in one CPSC block, we can further obtain
\begin{equation}\label{equ:sys_model34} 
 \begin{array}{l}
{\bf{y}}_{}^{ } = {\bf{\tilde H}}_{}^{ }{\bf{x}}_{}^{ } + {\bf{w}}^{ },
   \end{array}
\end{equation}
where ${\bf{y}}_{}^{ } = {\left[ {{{\left( {{\bf{y}}_1^{ }} \right)}^{\rm{T}}}{{\left( {{\bf{y}}_2^{ }} \right)}^{\rm{T}}} \cdots {{\left( {{\bf{y}}_Q^{ }} \right)}^{\rm{T}}}} \right]^{\rm{T}}} \in {^{\left( {{M_{{\rm{RF}}}}Q} \right)}}$, the aggregate SM signal ${\bf{x}}_{}^{ } = {\left[ {{{\left( {{\bf{x}}_1^{ }} \right)}^{\rm{T}}}{{\left( {{\bf{x}}_2^{ }} \right)}^{\rm{T}}} \cdots {{\left( {{\bf{x}}_Q^{ }} \right)}^{\rm{T}}}} \right]^{\rm{T}}} \in {^{\left( {K{n_t}Q} \right)}}$, ${\bf{w}}_{}^{ } = {\left[ {{{\left( {{\bf{w}}_1^{ }} \right)}^{\rm{T}}}{{\left( {{\bf{w}}_2^{ }} \right)}^{\rm{T}}} \cdots {{\left( {{\bf{w}}_Q^{ }} \right)}^{\rm{T}}}} \right]^{\rm{T}}}$, and
\begin{equation}
\begin{small}
 \begin{array}{l}
{\bf{\tilde H}}_{}^{^{ }} = \left[ {\begin{array}{*{20}{c}}
{{\bf{\tilde H}}_0^{^{ }}}&{\bf{0}}&{\bf{0}}& \cdots &{{\bf{\tilde H}}_2^{^{ }}}&{{\bf{\tilde H}}_1^{^{ }}}\\
{{\bf{\tilde H}}_1^{^{ }}}&{{\bf{\tilde H}}_0^{^{ }}}&{\bf{0}}& \cdots & \vdots &{{\bf{\tilde H}}_2^{^{ }}}\\
 \vdots &{{\bf{\tilde H}}_1^{^{ }}}&{{\bf{\tilde H}}_0^{^{ }}}& \cdots &{{\bf{\tilde H}}_{P - 1}^{^{ }}}& \vdots \\
{{\bf{\tilde H}}_{P - 1}^{^{ }}}& \vdots &{{\bf{\tilde H}}_1^{^{ }}}& \cdots &{\bf{0}}&{{\bf{\tilde H}}_{P - 1}^{^{ }}}\\
{\bf{0}}&{{\bf{\tilde H}}_{P - 1}^{^{ }}}& \vdots & \vdots & \vdots &{\bf{0}}\\
 \vdots &{\bf{0}}&{{\bf{\tilde H}}_{P - 1}^{^{ }}}& \vdots & \vdots & \vdots \\
 \vdots & \vdots &{\bf{0}}& \vdots &{\bf{0}}& \vdots \\
 \vdots & \vdots & \vdots & \vdots &{{\bf{\tilde H}}_0^{^{ }}}&{\bf{0}}\\
{\bf{0}}&{\bf{0}}&{\bf{0}}& \cdots &{{\bf{\tilde H}}_1^{^{ }}}&{{\bf{\tilde H}}_0^{^{ }}}
\end{array}} \right].
 \end{array}
 \end{small}
\end{equation}
 The SNR at the receiver is defined by $\mbox{SNR}={\rm E}\{ \|
 \mathbf{\tilde H}^{ } \, \mathbf{x}^{ }\|_2^2\} / {\rm E}\{\|\mathbf{w}^{ }\|_2^2\}$.

To detect the aggregate SM signal $\mathbf{x}$ from
 (\ref{equ:sys_model34}), the optimal signal detector relies on
 the ML algorithm:
\begin{equation}\label{equ:sys_model4} 
\begin{array}{l}
\mathop {\min }\limits_{{{{\bf{\hat x}}}^{ }}} {\left\| {{{\bf{y}}^{ }} - {{{\bf{\tilde H}}}^{ }}{{{\bf{\hat x}}}^{ }}} \right\|_2} = \mathop {\min }\limits_{\{ {\bf{\hat x}}_{k,q}^{ }\} _{k = 1,q = 1}^{K,Q}} {\left\| {{{\bf{y}}^{ }} - {{{\bf{\tilde H}}}^{ }}{{{\bf{\hat x}}}^{ }}} \right\|_2}, \\
 {\rm s.t.} ~ {\rm supp}\left( {{\bf{\hat x}}_{k,q}^{ }} \right) \in \mathbb{A} , \,
 \left.{{\bf{\hat x}}_{k,q}^{ }} \right\rangle_{{\rm supp}\left( {{\bf{\hat x}}_{k,q}^{ }} \right)}
 \in \mathbb{L} , \\ \, 1\le k\le K ,1\le q\le Q ,
 \end{array}
\end{equation}
whose complexity increases exponentially with the number of users,
since the size of the search set for the ML detector is $(n_t\cdot
L)^{KQ}$. This excessive complexity can be unaffordable in practice. To reduce the complexity, near-optimal sphere decoding detectors have
been proposed~\cite{SD}, but their complexity may still remain high,
particularly for the systems supporting large $K$, $Q$, $n_t$, and $L$~\cite{CS_CL1}. 
In conventional LS-MIMO
systems, low-complexity linear signal detectors (e.g., the MMSE-based signal detector)
have been shown to be near-optimal since $M=M_{\rm{RF}}\gg K$ and $n_t=1$ lead the
multi-user signal detection to be an \emph{over-determined} problem~\cite{LS_MIMO}.  
However, in the proposed
scheme, 
we have $M_{\rm{RF}}< Kn_t$. Hence the multi-user signal detection
problem (\ref{equ:sys_model4}) represents a large-scale
\emph{under-determined} problem. Consequently, the conventional linear
signal detectors perform poorly in the proposed LS-SM-MIMO \cite{CS_CL1}. By
exploiting the sparsity of SM signals, the authors
of~\cite{CS_CL1,CS_CL2,{shim}} have proposed the concept of CS-based
signal detectors for the downlink of small-scale SM-MIMO operating in
a single-user scenario. However, these signal detectors are unsuitable for the proposed multi-user
scenarios.
Observe from (\ref{equ:sys_model0}) that ${{\bf{ x}}_{k,q}^{ }}$ is a sparse
signal having a sparsity level of one. Hence the aggregate SM signal $\mathbf{x}^{ }$ which
consists of multiple users' SM signals in $Q$ time slots exhibits the distributed sparsity with the sparsity
level of $KQ$. This property of $\mathbf{x}^{}$ inspires us to exploit
SCS theory for the multi-user signal detection \cite{STR_CS}.  To further improve the signal detection performance and to
increase the system's throughput, we propose a joint SM transmission
scheme and an SCS-based MUD, which will be detailed in the next
section.

\section{SCS-Based MUD for LS-SM-MIMO UL}\label{S3}

To solve the multi-user signal detection of our UL LS-SM-MIMO system, we first propose
a joint SM transmission scheme to be employed at the users. Accordingly, an
SCS-based low-complexity MUD is developed at the BS, whereby
the distributed sparsity of the aggregate SM signal and
the group sparsity of multiple aggregate SM signals are
exploited. Moreover, the computational complexity of the proposed SCS-based MUD is discussed.

\subsection{Joint SM Transmission Scheme at the Users}\label{S3.1}

For the $k$th user in the $q$th time slot, every successive $J$ CPSC block are considered as a group and share the
same spatial constellation symbol, namely,
\begin{equation}\label{commonsparse} 
  \begin{array}{l}
 {\rm supp}\left( \mathbf{x}_{k,q}^{(1)} \right) = {\rm supp}\left( \mathbf{x}_{k,q}^{(2)} \right)
 =  \cdots  = {\rm supp}\left( \mathbf{x}_{k,q}^{(J)} \right) ,\\1\le k \le K, 1\le q \le Q,
   \end{array}
\end{equation}
 where we introduce the superscript $(j)$ to denote the $j$th CPSC block, and $J$ is typically small, e.g.,
 $J=2$. In CS theory, the specific signal structure, where $\mathbf{x}_{k,q}^{(1)},
 \mathbf{x}_{k,q}^{(2)},\cdots , \mathbf{x}_{k,q}^{(J)}$ share a common
 support is often referred to as \emph{group sparsity}. Similarly, the
 aggregate SM signals 
 also have the group sparsity, i.e.,
\begin{equation}\label{commonsparse2} 
 {\rm supp}\left( \mathbf{x}^{(1)} \right) = {\rm supp}\left( \mathbf{x}^{(2)} \right)
 =  \cdots  = {\rm supp}\left( \mathbf{x}^{(J)} \right) ,
\end{equation}
 Although exhibiting group sparsity may slightly reduce the
 information carried by the spatial constellation symbols, it is also
 capable of reducing the number of the RF chains required according to
 the SCS theory, whilst simultaneously improving the total bit error
 rate (BER) of the entire system even with higher UL-throughput. This conclusion will be 
 confirmed by our simulation results. 

\subsection{SCS-Based MUD at the BS}\label{S3.2}

According to~(\ref{equ:sys_model34}), the received signals at the BS in the same group can be expressed~as
\begin{equation}\label{equ:sys_model3} 
  \begin{array}{l}
 \mathbf{y}^{(j)} = \mathbf{\tilde H}^{(j)} \mathbf{ x}^{(j)} + \mathbf{w}^{(j)} , ~ 1\le j \le J ,
   \end{array}
\end{equation}
 where $\mathbf{y}^{(j)}$ denotes the received signal in the $j$th
 CPSC block, while $\mathbf{\tilde H}^{(j)}$ and $\mathbf{w}^{(j)}$ are the
 effective MIMO channel matrix and the AWGN vector,
 respectively.

 The intrinsically distributed sparsity of $\mathbf{x}^{(j)}$ and the
 under-determined nature of~(\ref{equ:sys_model3}) inspire us to solve
 the signal detection problem based on CS theory, which can efficiently acquire the sparse solutions to under-determined linear systems. Moreover, the $J$ different aggregate SM signals in
 (\ref{equ:sys_model3}) can be jointly exploited for improving the
 signal detection performance due to the group sparsity of
 $\{ {\mathbf{x}}^{(j)} \}_{j=1}^{J}$. Thus, by considering both the
 distributed sparsity and the group sparsity of the aggregate SM signals, the multi-user signal detection at the BS can be formulated as the following
 optimization problem
\begin{equation}\label{equ:sys_model5} 
\begin{array}{l}
\!\!\! \min\limits_{\{\widehat{\mathbf{x}}^{(j)}\}_{j=1}^J}\!\! \sum\limits_{j= 1}^J \! \left\| \mathbf{y}^{(j)}
 \! - \! \mathbf{\tilde H}^{(j)} \widehat{\mathbf{x}}^{(j)} \right\|_2 ^2 \!\!\\ = \!\!
 \min\limits_{\{\widehat{\mathbf{x}}_{k,q}^{(j)}\}_{j=1,k=1,q=1}^{J,K,Q}}\!\!\sum\limits_{j= 1}^J \! \left\| \mathbf{y}^{(j)}
 \! - \! \mathbf{\tilde H}^{(j)} \widehat{\mathbf{x}}^{(j)} \right\|_2 ^2,  \\
 \hspace*{6mm}{\rm s.t.} ~ \left\| \widehat{\mathbf{x}}_{k,q}^{(j)} \right\|_0 = 1 , ~ 1\le j\le J,~ 1\le q\le Q, ~
 1\le k\le K .
\end{array}
\end{equation}
 Our proposed SCS-based MUD solves the optimization problem
 (\ref{equ:sys_model5}) with the aid of two steps. In the first step,
 we estimate the spatial constellation symbols, i.e., the indices of
 $K$ users' active AEs in $J$ successive CPSC blocks. In the
 second step, we infer the legitimate signal constellation symbols of
 the $K$ users in $J$ CPSC blocks.

\subsubsection{Step 1. Estimation of Spatial Constellation Symbols}

 We propose a group subspace pursuit (GSP) algorithm developed
 from the classical subspace pursuit (SP) algorithm of~\cite{SP} to
 acquire the sparse solution to the large-scale under-determined
 problem (\ref{equ:sys_model5}), where both the {\em apriori} sparse
 information (i.e., $\left\|\mathbf{x}_{k,q}^{(j)}\right\|_0=1$) and the
 group sparsity of $\mathbf{x}^{(1)},\mathbf{x}^{(2)},\cdots
 ,\mathbf{x}^{(J)}$ are exploited for improving the multi-user signal detection
 performance. The proposed GSP algorithm is described in
 \textbf{Algorithm~\ref{alg:Framwork}}, which estimates SM
 signal $\left\{ {\widehat {\bf{x}}_{k,q}^{(j)}} \right\}_{k = 1,j = 1,q = 1}^{K,J,Q}$. Hence, the estimated spatial constellation
 symbol is $\left\{ {{\rm{supp}}\left( {\widehat {\bf{x}}_{k,q}^{(j)}} \right)} \right\}_{k = 1,j = 1,q = 1}^{K,J,Q}$.

\begin{algorithm}[bp!]
\begin{small}
\renewcommand{\algorithmicrequire}{\textbf{Input:}}
\renewcommand\algorithmicensure {\textbf{Output:} }
\caption{Proposed GSP Algorithm.}
\label{alg:Framwork} 
\begin{algorithmic}[1]
\REQUIRE
 Noisy received signals $\mathbf{y}^{(j)}$ and effective channel matrices $\mathbf{\tilde H}^{(j)}$
 for $1\le j\le J$.
\ENSURE
 Estimated $\widehat{\mathbf{x}}^{(j)} = \left[
 \left(\widehat{\mathbf{x}}_1^{(j)}\right)^{\rm T} ~ \left(\widehat{\mathbf{x}}_2^{(j)}\right)^{\rm T}
 \cdots \left(\widehat{\mathbf{x}}_Q^{(j)}\right)^{\rm T} \right]^{\rm T}$, where $\widehat{\mathbf{x}}_q^{(j)} = \left[
 \left(\widehat{\mathbf{x}}_{1,q}^{(j)}\right)^{\rm T} ~ \left(\widehat{\mathbf{x}}_{2,q}^{(j)}\right)^{\rm T}
 \cdots \left(\widehat{\mathbf{x}}_{K,q}^{(j)}\right)^{\rm T} \right]^{\rm T}$ for $1\le q \le Q$. \\
\STATE $\mathbf{r}^{(j)}=\mathbf{y}^{(j)}$ for $1\le j\le J$;~\{Initialization\}
\STATE $\Omega^0 = \emptyset$;  ~\{Empty support set\}
\STATE $t = 1$; ~\{Iteration index\}
\REPEAT
 \STATE $\mathbf{a}_{k,q}^{(j)}=\left(\mathbf{\tilde H}_{k,q}^{(j)}\right)^{*} \mathbf{r}^{(j)}$ for
  $1\le k \le K$, $1\le q \le Q$, and $1\le j\le J$;~ \{Correlation\}
 \STATE $\tau_{k,q}=\arg \max\limits_{\widetilde{\tau}_{k,q}} \sum\limits_{j=1}^J \left\| \left.
  \mathbf{a}_{k,q}^{(j)} \right\rangle_{\widetilde{\tau}_{k,q}} \right\|_2^2$ for $1\le k \le K$, $1\le q \le Q$;
~\{Identify support\}
 \STATE $\Gamma  = \left\{ {{\tau _{k,q}} + \left( {k - 1 + K\left( {q - 1} \right)} \right){n_t}} \right\}_{k = 1,q = 1}^{K,Q}$;
 ~\{Preliminary support set\}
 \STATE $\left. \mathbf{b}^{(j)} \right|_{\Omega^{t-1}\cup \Gamma} = \left( \left. \mathbf{\tilde H}^{(j)} \right|_{\Omega^{t-1}\cup \Gamma} \right)^\dag
  \mathbf{y}^{(j)}$ for $1\le j\le J$; ~\{Least squares\}
 \STATE $\omega_{k,q} = \arg \max\limits_{\widetilde{\omega}_{k,q}} \sum\limits_{j=1}^J \left\|
  \left. \mathbf{b}_{k,q}^{(j)} \right\rangle_{\widetilde{\omega}_{k,q}} \right\|_2^2$ for $1\le k \le K$, $1\le q \le Q$;
~ \{Pruning support set\}
 \STATE $\Omega^t = \left\{ {{\omega _{k,q}} + \left( {k - 1 + K\left( {q - 1} \right)} \right){n_t}} \right\}_{k = 1,q = 1}^{K,Q}$; ~\{Final support set\}
 \STATE $\left. \mathbf{c}^{(j)} \right|_{\Omega^t} = \left( \left. \mathbf{\tilde H}^{(j)}
  \right|_{\Omega^t} \right)^\dag \mathbf{y}^{(j)}$ for $1\le j\le J$; ~~${\kern -1.0pt} $~\{Least squares\}
 \STATE $\mathbf{r}^{(j)} = \mathbf{y}^{(j)} - \mathbf{\tilde H}^{(j)} \mathbf{c}^{(j)}$ for
  $1\le j\le J$; ~ \{Compute residual\}
   \STATE $t = t+1$;~\{Update iteration index\}
\UNTIL{$\Omega^t=\Omega^{t-1}$ or $t\ge Q$}
\end{algorithmic}
\end{small}
\end{algorithm}

 Compared to the classical SP algorithm, the proposed GSP
 algorithm exploits the distributed sparsity and group sparsity of $\left\{ {{{\bf{x}}^{\left( j \right)}}} \right\}_{j = 1}^J$. More explicitly,
 $\mathbf{x}^{(j)}\in \mathbb{C}^{(KQn_t)}$ consists of the $KQ$ low-dimensional sparse vectors
 $\mathbf{x}_{k,q}^{(j)}\in \mathbb{C}^{n_t}$, where each $\mathbf{x}_{k,q}^{(j)}$ has
 the known sparsity level of one, and the
 aggregate SM signals $\mathbf{x}^{(1)}, \mathbf{x}^{(2)}, \cdots ,
 \mathbf{x}^{(J)}$ appear the group sparsity. Specifically, the differences between the
 proposed GSP algorithm and the classical SP algorithm lie in the
 following two aspects: 1)~the identification of support set including the steps of \emph{preliminary support set} and \emph{final support set} as shown in
 \textbf{Algorithm~\ref{alg:Framwork}}; and 2)~the joint processing
 of $\mathbf{y}^{(1)}, \mathbf{y}^{(2)}, \cdots ,
 \mathbf{y}^{(J)}$. First, for the support selection, taking the step of \emph{preliminary support set} for instance, when selecting the preliminary
 support set, the classical SP algorithm selects the support set associated with the first $KQ$ largest values of the global correlation result
 $\left( \mathbf{\tilde H}^{(j)} \right)^{*} \mathbf{r}^{(j)}$. By contrast,
 the proposed GSP algorithm selects the support set associated with the
 largest value from the local correlation result in each
 $\left(\mathbf{\tilde H}_{k,q}^{(j)}\right)^{*}\mathbf{r}^{(j)}$.  In this way, the distributed sparsity of the aggregate SM signal can be exploited for improved signal detection performance. 
Second, compared with the classical SP algorithm, the proposed GSP algorithm jointly exploits the $J$
 correlated signals having the group sparsity, which can bring the further improved signal detection performance.

 It should be noted that even for
 the special case of $J=1$, i.e., without using the joint SM
 transmission scheme, the proposed GSP algorithm still achieves a
 better signal detection performance than the classical SP algorithm
 when handling the aggregate SM signal, since the inherently
 distributed sparsity of the aggregate SM signal is leveraged to
 improve the signal detection performance.

\subsubsection{Step 2. Acquisition of Signal Constellation Symbols}
 Following \textit{Step 1}, we can also acquire a rough estimate of
 the signal constellation symbol for each user in each time slot. By searching for the
 minimum Euclidean distance between this rough estimate of the
 signal constellation symbol and the legitimate constellation symbols of
 $\mathbb{L}$, we can obtain the final estimate of signal constellation symbols.

\subsection{Computational Complexity}\label{S3.3}
The optimal ML signal detector has a prohibitively high
 computational complexity of ${\cal O}\big( (L\cdot n_t)^{(K \cdot Q)}\big)$
 according to~(\ref{equ:sys_model4}). 
 The sphere decoding detectors \cite{SD} are indeed capable of
 reducing the computational complexity, but they may still suffer from
 an unaffordable complexity, particularly for large $K$, $Q$, $L$ and $n_t$
 values. By contrast, the conventional MMSE-based detector for LS-MIMO and CS-based detector \cite{CS_CL2} for small-scale SM-MIMO enjoy the low complexity of ${\cal O}\big(M_{\rm{RF}} \cdot (n_t \cdot
  Q\cdot K)^2+(n_t \cdot Q\cdot K)^3\big)$ and ${\cal O}\big(2M_{\rm{RF}} \cdot (Q\cdot K)^2 +
 (Q\cdot K)^3)$, respectively.
 For the proposed SCS-based MUD, most of the computational
 requirements are imposed by the least squares (LS) operations, which has a
 complexity of ${\cal O}\big(J\cdot (2M_{\rm{RF}} \cdot
(Q\cdot K)^2 + (Q\cdot K)^3) \big)$ \cite{LS}.  Consequently, the computational
 complexity per CPSC block is ${\cal O}\big(2M_{\rm{RF}} \cdot (Q\cdot K)^2 +
 (Q\cdot K)^3)$, since $J$ successive aggregate SM signals are jointly processed.
 Compared with conventional signal detectors, the proposed SCS-based MUD benefits from a substantially lower
 complexity than the ML signal detector or the sphere decoding detectors, and it
 has a similar low complexity to the conventional MMSE-based and CS-based signal detectors.



\section{Simulation Results}\label{S4}
\begin{figure}[tp]
\setlength{\abovecaptionskip}{-10pt}
\setlength{\belowcaptionskip}{0pt}
\begin{center}
\includegraphics[width=9cm]{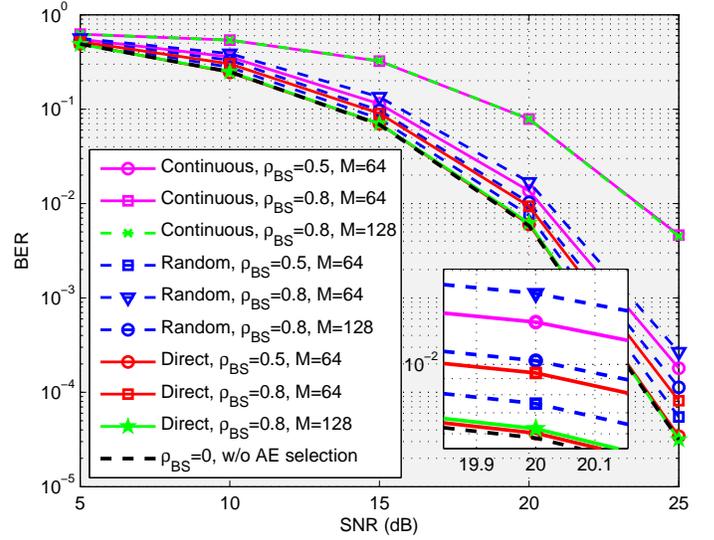}
\end{center}
\caption{The total BERs achieved by the proposed SCS-based MUD with different AE selection schemes, where $K=8$, $J=2$, 64QAM, $M_{\rm{RF}}=18$, $n_t=4$, and $\rho_{\rm{US}}=0$ are considered.}
\label{FIG2}
\end{figure}
A simulation study was carried out to compare the performance of the
 proposed SCS-based MUD with that of the MMSE-based signal detector~\cite{LS_MIMO}
 and the CS-based signal detector~\cite{CS_CL2}.
 In the LS-SM-MIMO system considered, the BS with ULA
 employed a large number of AEs $M$ but a much smaller number of RF chains $M_{\rm{RF}}$,
 while $K$ users employing $n_t$ AEs but only a single RF chain simultaneously use CPSC scheme with $P=8$ and $Q=64$ to transmit SM signals to the BS. The total BER including the spatial constellation
 symbol and the signal constellation symbol were investigated. 

\begin{figure}[tp]
\setlength{\abovecaptionskip}{-10pt}
\setlength{\belowcaptionskip}{0pt}
\begin{center}
\includegraphics[width=9cm]{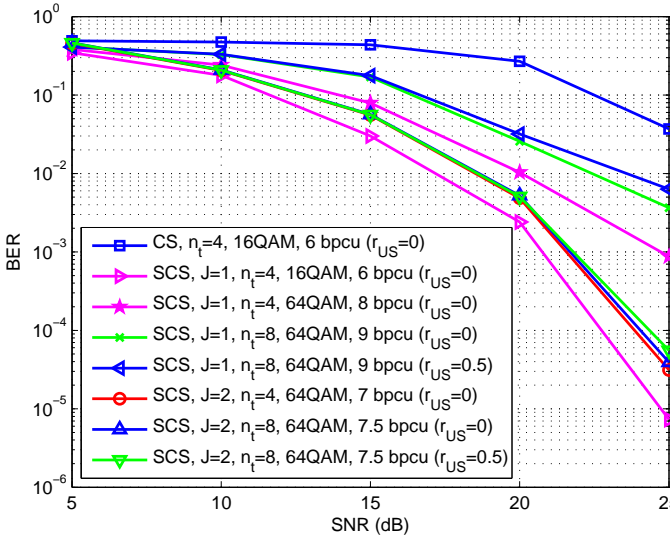}
\end{center}
\caption{The total BERs achieved by CS-based signal detector and SCS-based MUD against different SNR's in LS-SM-MIMO, where $K=8$, $M_{\rm{RF}}=18$, $M=64$, $\rho_{\rm{BS}}=0.5$, and the direct AE selection scheme is considered.}
\label{FIG3}
\end{figure}
 Fig.~\ref{FIG2} compares the total BERs achieved by the proposed SCS-based MUD with different AE selection schemes, where $K=8$, $J=2$, 64QAM, $M_{\rm{RF}}=18$, $n_t=4$, and $\rho_{\rm{US}}=0$ are considered. The continuous AE selection scheme means to select $M_{\rm{RF}}$ continuous AEs, i.e., $\Theta  = \left\{ {\varphi  + {m_{{\rm{RF}}}}} \right\}_{{m_{{\rm{RF}}}} = 0}^{{M_{{\rm{RF}}}} - 1}$ with $1\le \varphi \le M-M_{\rm{RF}}+1 $. The random AE selection scheme means that elements of $\Theta $ are uniquely selected from the set $\left\{ {1,2, \cdots ,M} \right\}$ randomly. The direct AE selection scheme \cite{mide} has been provided in Section \ref{S2.2}. Besides, the BER achieved by the SCS-based MUD with $\rho_{\rm{BS}}=0$ is considered as the performance bound, since $\rho_{\rm{BS}}=0$ and $\rho_{\rm{US}}=0$ imply the uncorrelated Rayleigh-fading MIMO channels. From Fig.~\ref{FIG2}, it can be observed that the direct AE selection scheme outperforms two other AE selection schemes. Moreover, for a certain AE selection scheme, the BER performance degrades when $M_{\rm{RF}}/M$ or $\rho_{\rm{BS}}$ increases. For the direct AE selection scheme, the BER performance with the case of $\rho_{\rm{BS}}=0.8$, $M=128$ and the case of $\rho_{\rm{BS}}=0.5$, $M=64$ approaches the BER achieved over uncorrelated Rayleigh-fading MIMO channels, which indicates the near-optimal performance of the direct AE selection scheme.


 Fig.~\ref{FIG3} compares the total BERs achieved by CS-based signal detector and the proposed SCS-based MUD against different SNR's in LS-SM-MIMO, where $K=8$, $M_{\rm{RF}}=18$, $M=64$, $\rho_{\rm{BS}}=0.5$, and the direct AE selection scheme is considered. The SCS-based MUD outperforms the CS-based signal detector even for $J=1$, since the distributed sparsity of the aggregate SM signal is exploited. For the SCS-based MUD, the BER performance improves when $J$ increases, at the cost of the reduced UL-throughput. To mitigate this issue, the larger number of AEs can be used at the users to achieve the higher modulation order of the spatial constellation symbol set. Specifically, by increasing $n_t$ from 4 to 8, the UL-throughput of the SCS-based MUD can increase, but more AEs at the user indicates the larger $\rho_{\rm{US}}$. When $n_t$ increases, the BER performance of the SCS-based MUD with $J=1$ degrades obviously. By contrast, when $n_t$ increases, the BER performance loss of the SCS-based MUD with $J=2$ can be nonnegligible, even when the larger $\rho_{\rm{US}}$ for the larger $n_t$ is considered.

\begin{figure}[tp]
\setlength{\abovecaptionskip}{-10pt}
\setlength{\belowcaptionskip}{0pt}
\begin{center}
\includegraphics[width=9cm]{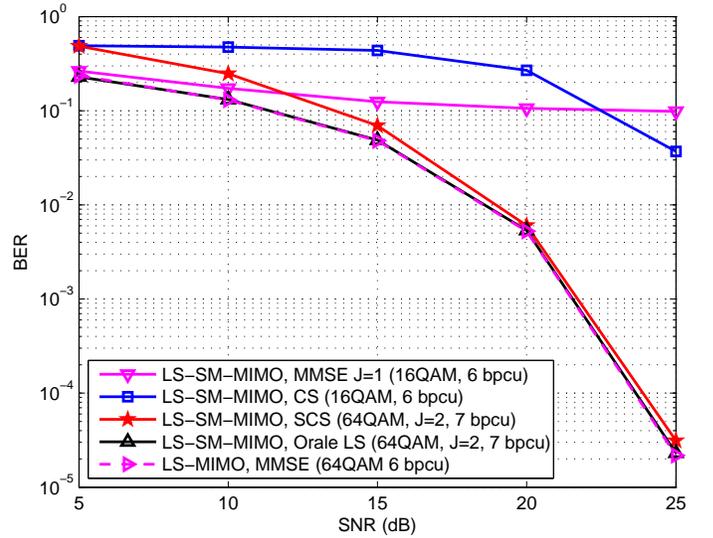}
\end{center}
\caption{The total BERs achieved by different signal detectors against different SNR's in the proposed LS-SM-MIMO and conventional LS-MIMO.}
\label{FIG4}
\end{figure}

Fig.~\ref{FIG4} depicts the total BERs achieved by different signal detectors against different SNR's in the proposed LS-SM-MIMO, where $K=8$, $M_{\rm{RF}}=18$, $M=64$, $n_t=4$, $\rho_{\rm{BS}}=0.5$, $\rho_{\rm{US}}=0$, and the direct AE selection scheme is considered.
In Fig.~\ref{FIG4}, we also provide the oracle LS-based signal detector with the known spatial constellation symbol perfectly known at the BS for the proposed LS-SM-MIMO with $J=2$, 64QAM and MMSE-based signal detector for LS-MIMO with 64QAM, where both of them only consider the BER of signal constellation symbol. Here we consider LS-MIMO uses the same number of RF chains to serve 8 single-antenna users over uncorrelated Rayleigh-fading channels.
The superior performance of our SCS-based MUD to the MMSE-based and CS-based signal detectors is clear. 
 Moreover, the performance gap between the oracle LS-based signal detector with 7 bpcu and the proposed SCS-based MUD with 7 bpcu is less than 0.5 dB. Note that the oracle LS-based signal detector only considers the BER of signal constellation symbol, while the proposed SCS-based MUD considers both the spatial and signal constellation symbols. Finally, compared with the conventional LS-MIMO with MMSE-based signal detector (6 bpcu), our proposed UL LS-SM-MIMO and the associated SCS-based MUD (7bpcu) only suffers from a negligible BER loss, which confirmed the improved UL-throughput of the proposed LS-SM-MIMO scheme.

\section{Conclusions}\label{S5}

 We have proposed a LS-SM-MIMO scheme for the UL-transmission. The BS employs a large
 number of AEs but a much smaller number of RF chains, where a simple receive AE selection scheme is used
 for the improved performance. Each user equipped with multiple AEs but only a single RF chain uses CPSC to combat multipath channels.
 SM has been adopted for the UL-transmission to improve the UL-throughput. The proposed scheme is especially suitable for scenarios,
 where a large number of low-cost AEs can be accommodated, and
 both the power consumption as well as hardware cost are heavily
 determined by the number of RF chains. Due to the reduced number of
 RF chains at the BS and multiple AEs employed by each user, the UL
 multi-user signal detection is a challenging large-scale under-determined
 problem. We have proposed a joint SM transmission scheme at the users to introduce the group sparsity of multiple aggregate SM
 signals, and a matching SCS-based MUD at the BS has been
 proposed to leverage the inherently distributed sparsity of the
 aggregate SM signal as well as the group sparsity of multiple
 aggregate SM signals for reliable multi-user signal detection performance. The
 proposed SCS-based MUD enjoys the low complexity,
 and our simulation results have demonstrated that it performs better than
 its conventional counterparts with even much higher UL-throughput.


\end{document}